\documentclass[a4paper,11pt]{article}
\usepackage{pos}
\usepackage{bm}

\title{Hybrid stochastic method for the tensor renormalization group}

\author*[a]{Hiroshi Ohki}
\author[a]{Erika Arai}
\author[b]{Masaaki Tomii}

\affiliation[a]{Department of Physics, Nara Women's University, Nara 630-8506, Japan}

\affiliation[b]{
Physics Department, University of Connecticut, Storrs, CT 06269-3046, USA}

\emailAdd{hohki@asuka.phys.nara-wu.ac.jp}

\abstract{
We propose a hybrid stochastic method for the tensor renormalization group
(TRG) approach.
TRG is known as a powerful tool to study the many-body systems and
quantum field theory on the lattice.
It is based on a low-rank approximation of the tensor using the
truncated singular value decomposition (SVD),
whose computational cost increases as the bond dimension increases, 
so that efficient cost reduction techniques are highly demanded.
We use noise vectors for the low-rank approximation with the truncated SVD, 
by which the truncation error is replaced with a statistical error due to noise, and an error estimation could be improved. 
We test this method in the classical Ising model and observe a better accuracy than TRG.   
We also discuss a cross contamination issue in a multiple use of the same noise vectors,
and to remove this systematic error we consider position-dependent noise vectors.}

\FullConference{
 The 38th International Symposium on Lattice Field Theory, LATTICE2021
  26th-30th July, 2021
  Zoom/Gather@Massachusetts Institute of Technology
}

\begin{document}
\maketitle

\section{Introduction}
The tensor network is a powerful tool to study the many-body systems and quantum field theory on the lattice. 
In a tensor network representation the partition functions 
can be obtained by contracting the tensor products.
One efficient and simple algorithm for a tensor contraction
is the tensor renormalization group (TRG) approach~\cite{Levin:2006jai}.
Since TRG is a fully deterministic algorithm, 
there is no sign problem and 
this method can also be applied to systems with a
complex action 
such as finite density QCD, the real-time dynamics 
that are not easily accessible by the Monte Carlo methods.

TRG is a very efficient numerical method of a coarse graining, 
but contracting the tensor products is still the most time-consuming part,
whose computational cost scales as e.g., $\mathcal{O}(D_{\rm cut}^6)$ in the 2-D Ising model,
where $D_{\rm cut}$ is the truncated bond dimension used in the singular value decomposition (SVD). 
The computational cost significantly increases as the bond dimension increases, 
so that efficient cost reduction techniques are highly demanded.
In addition, there exists a systematic error, which arises from the truncation in the SVD.
Since an estimation of systematic errors becomes difficult 
when the system becomes more complicated, 
it is important to investigate a possibility of an alternative approach to improve the error estimation
that can be easily implemented on tensor networks.
It is also known that a short-distance effect remains after coarse graining~\cite{Gu:2009}, 
and the resulting coarse-grained tensor does not obey a correct RG transformation. 
To obtain a physically correct RG flow and to reduce the truncation error, 
other renormalization methods have been proposed, 
such as Loop-TNR~\cite{Yang:2017}, TNR~\cite{Evenbly:2015}, GILT~\cite{Hauru:2018}.
Also several truncation techniques such as a randomized SVD~\cite{Morita:2018}, a projective truncation method~\cite{Nakamura:2018enp}, 
have been proposed, which can reduce the computation cost as $\mathcal{O}(D_{\rm cut}^5)$.
We note, however, that a low-rank approximation of the tensor decompositions is commonly 
used in these methods, so that there remains a systematic uncertainty due to the truncation error.   

In this work, we propose a new approach based on a stochastic calculation
of the low-rank approximation of a tensor decomposition. 
We use random noise vectors in combining with the truncated SVD, 
in which the truncated bond dimension is given as the sum of the number of singular values and the noise vectors.
The truncation error can be replaced with an error due to the noise, 
and a stochastic determination of physical quantities is possible.  
A feasibility study of this method is shown for the 2-D Ising model, 
where we employ this new truncation method in the same coarse graining process 
as in TRG.  
We study the noise structure dependence of the free energy 
and a noise contamination issue which is found to be the only source of the systematic uncertainty.
We also discuss a possibility to completely remove the systematic error by employing position-dependent noise vectors.

\section{Rank truncation with random noise vector}
Let us start by briefly explaining the TRG method for the 2-D Ising model. 
The partition function $Z$ is given as a trace over the tensor indices in a tensor-network representation,  
\begin{align}
Z= \sum_{\{\sigma\}} e^{-\beta H} = \sum_{ijk\cdots}T_{ijkl}T_{mni\ell}\cdots = {\rm Tr}\left[\otimes T \right], 
\end{align}
where $T_{ijkl}$ is the initial tensor. 
We express the tensor in a matrix representation, $T_{ijkl} = M_{ij;kl} \equiv M_{ab}$, and $T_{ijkl} = \tilde{M}_{jk;li} \equiv \tilde{M}_{cd}$. 
Using the SVD, we represent a matrix $M$ with a rank $R$ as follows,
\begin{align}
M_{ab} = \sum_{s=1}^R \sqrt{\Lambda_s} u_{as} \sqrt{\Lambda_s} v_{sb}, 
\end{align}
where $\Lambda_s$ is a singular value ($\Lambda_1 \geq \Lambda_2 \geq \Lambda_3 \cdots$).
Then the fourth-order tensor is approximated by a product of two low-rank third-order tensors with a rank $D \leq R$ 
via the SVD,
\begin{align}
\label{eq:SVD}
T_{ijkl} = M_{ab} \simeq \sum_{s=1}^D \sqrt{\Lambda_s} u_{as} \sqrt{\Lambda_s} v_{sb} = 
\sum_{s=1}^D S_{3as}S_{1sb},
\end{align}
where $S_{3as}=\sqrt{\Lambda_s} u_{as}$, and $S_{1sb}=\sqrt{\Lambda_s} v_{sb}$ 
are the third-order tensors. 
Similarly, $S_{2,4}$ are also defined to approximate the other type of a tensor decomposition, 
$\tilde{M}_{cd} \simeq \sum_{s=1}^D S_{4cs} S_{2sd}$.
We then consider a coarse graining by contracting 
all the internal indices of the four third-order tensors of $S_{1,2,3,4}$. 
By repeating these processes $n$-times we obtain a renormalized tensor $T^{(n)}_{i'j'k'l'}$ as
\begin{align}
T^{(n)}_{i'j'k'l'} = \sum_{ijkl} S^{(n-1)}_{1i';ij}S^{(n-1)}_{2j';kj} S^{(n-1)}_{3lk;k'} S^{(n-1)}_{4il;l'}
\equiv {\rm Tr}[ S^{(n-1)}_{1i'}S^{(n-1)}_{2j'} S^{(n-1)}_{3k'} S^{(n-1)}_{4l'}], \ \ (n >0)
\end{align}
where $T^{(0)}=T$ and $S^{(0)}_{1,2,3,4}=S_{1,2,3,4}$.
The bond dimension of the renormalized tensor $D_{\rm cut}$ is determined by the number of the 
singular values which also determines the accuracy of the method.  
The dominant computational cost for contractions scales as $\mathcal{O}(D_{\rm cut}^6)$, 
(the randomized TRG $\mathcal{O}(D_{\rm cut}^5))$.

\begin{figure}[tbph]
\begin{center}
\includegraphics[clip,width=0.25\textwidth]{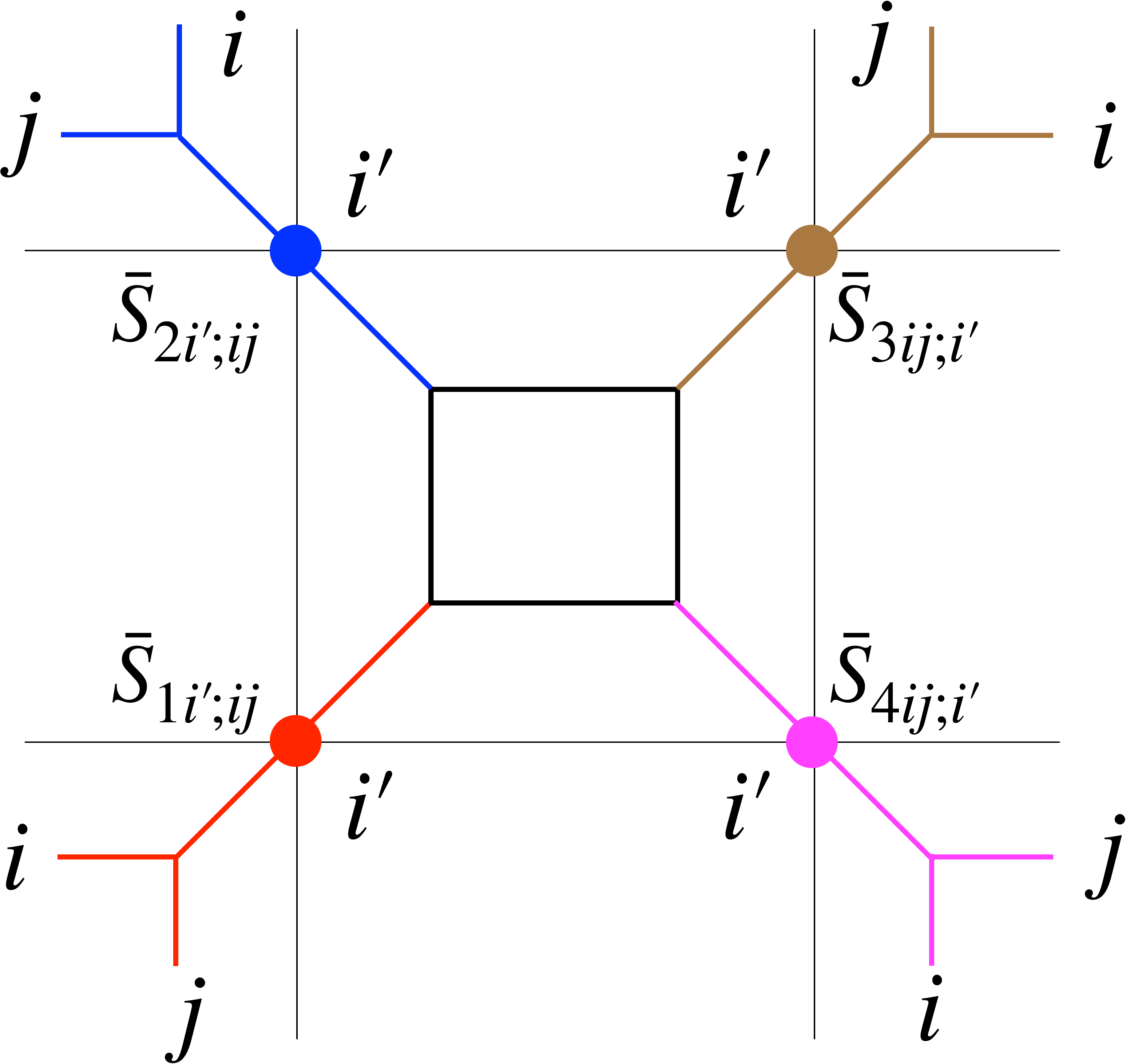} 
\end{center}
\caption{\label{fig:TRG} 
Tensor decomposition and modified third order tensors in the random noise vector method.}
\end{figure}

As shown above, TRG and other TRG-inspired renormalization methods
use a low-rank approximation technique to reduce the computational cost, 
while it becomes a source of the systematic uncertainty.  
In order to reduce the systematic uncertainty with a limited number of bond dimensions, 
we propose a different rank truncation method using random noises.
We introduce $D$-dimensional noise vectors $\eta_r=(\eta_{1r}, \eta_{2r}, \cdots, \eta_{Dr} )^T$ for $r=1,\cdots, N_r$, 
where each vector component $\eta_{ir}$ has a random number.
We define a $D \times N_r$ matrix $\bm{\eta} = (\eta_1, \cdots, \eta_{N_r})$,
which satisfies the completeness condition 
\begin{align}
\label{eq:noise}
\frac{1}{N_r} {\bm\eta} \cdot {\bm\eta}^\dagger = \sum_{r=1}^{N_r} \frac{1}{N_r} \eta_r \cdot \eta_r^\dagger =
{\bm 1}_{D\times D} + \mathcal{O}(1/\sqrt{N_r}).
\end{align}
This random noise vector can be used to decompose a matrix in combing with the SVD.
Substituting ${\bm 1} \sim \frac{1}{N_r}{\bm \eta}\cdot{\bm \eta}^\dagger$ into Eq.~\eqref{eq:SVD} 
with $D=R-D_{\rm svd}$ we approximate the matrix $M$ via the SVD as 
\begin{align}
M_{ab} 
&\simeq \sum_{s=1}^{D_{\rm svd}} (\sqrt{\Lambda_s} u_{as} \sqrt{\Lambda_s} v_{sb}) 
+ \sum_{s,t=D_{\rm svd}+1}^R 
\sqrt{\Lambda_s} u_{as} \left(\frac{1}{N_r}{\bm \eta}\cdot{\bm \eta}^\dagger\right)_{st} \sqrt{\Lambda_t} v_{tb}.
\end{align}
From this decomposition we suggest a modification of the third-order tensor to include the noise vector part.
For this purpose, we introduce the following modified third-order tensors as a function of the noise vectors, 
\begin{align}
\bar{S}_{3as}({\bm \eta}) &\equiv 
\begin{cases}
\sqrt{\Lambda_s} u_{as} & (1 \leq s \leq D_{\rm svd}) \\
\sum_{i=D_{\rm svd}+1}^R \sqrt{\frac{\Lambda_i}{N_r}} u_{ai} \eta_{is} & (D_{\rm svd}+1 \leq s \leq D_{\rm svd}+N_r)
\end{cases},
\notag \\
\bar{S}_{1sb}({\bm \eta}) &\equiv 
\begin{cases}
\sqrt{\Lambda_s} v_{sb} & (1 \leq s \leq D_{\rm svd}) \\
\sum_{i=D_{\rm svd}+1}^R \sqrt{\frac{\Lambda_i}{N_r}} \eta_{si}^* v_{ib} & (D_{\rm svd}+1 \leq s \leq D_{\rm svd}+N_r)
\end{cases}, 
\end{align}
and $\bar{S}_{2,4}({\bm \eta})$ are also defined similarly.
Using the modified tensors $\bar{S}_{1,2,3,4}$ with Eq.~\eqref{eq:noise} we immediately see 
$\displaystyle M_{ab} = \lim_{N_r \to \infty} \sum_{s=1}^{D_{\rm svd} + N_r} \bar{S}_{3as} \bar{S}_{1sb}$ 
for any value of $D_{\rm svd}$. 
Therefore, we can stochastically calculate the matrix decomposition.
Since the modified third-order tensors include all the modes for the singular values, 
there is only a statistical error due to noise vectors instead of the truncation error, 
which ensures that the original matrix should be reproduced within the statistical error. 
It should be emphasized that a similar decomposition has already been employed in lattice QCD calculations
for solving the Dirac equation, 
where a dominant part of the Dirac propagator is obtained using a low-mode part of the Dirac eigenvectors, 
and a high-mode part is stochastically evaluated using the random noise vectors.
The graphical representation of the tensor decomposition is given in Fig.~\ref{fig:TRG}.

\section{Hybrid stochastic method}

While the truncation error can be replaced with the statistical error, 
the bond dimension increases as $N_r$ increases
and the low-rank approximation of the matrix decomposition is not as good as the original TRG 
in the sense of the Frobenius norm due to the theorem by Eckhart-Young-Mirsky. 
Instead of increasing $N_r$, 
we adopt a statistical approach. 
Namely, we generate an ensemble of random noise vectors 
${\bm\eta}^{[\ell]} \ (\ell=1,\cdots,N)$ with $N$ being the total number of statistics, 
and $\ell$ labels a sample number. 
Our strategy is that 
a matrix decomposition is approximately obtained from an ensemble average of the 
modified third-order tensors with keeping a lower value of $N_r$. 
In fact, in this hybrid stochastic method
we can decompose 
the matrix as follows, 
\begin{align}
\displaystyle M_{ab} = \frac{1}{N} \sum_{\ell=1}^N \sum_{s=1}^{D_{\rm svd}+N_r} \bar{S}_{3as} ({\bm \eta}^{[\ell]}) \bar{S}_{1sb} ({\bm \eta}^{[\ell]})
+ \mathcal{O}(1/\sqrt{N_rN}).
\end{align}
Thus, an exact matrix decomposition can be obtained in the infinite statistical limit ($N \to \infty$)
while keeping $N_r$ finite.
This property is used to effectively reduce the bond dimensions.
In the next section, we discuss how to obtain a renormalized tensor using the above 
tensor decompositions.

\subsection{Coarse graining with common noise}

\begin{figure}[tbph]
\begin{center}
\includegraphics[clip,width=0.6\textwidth]{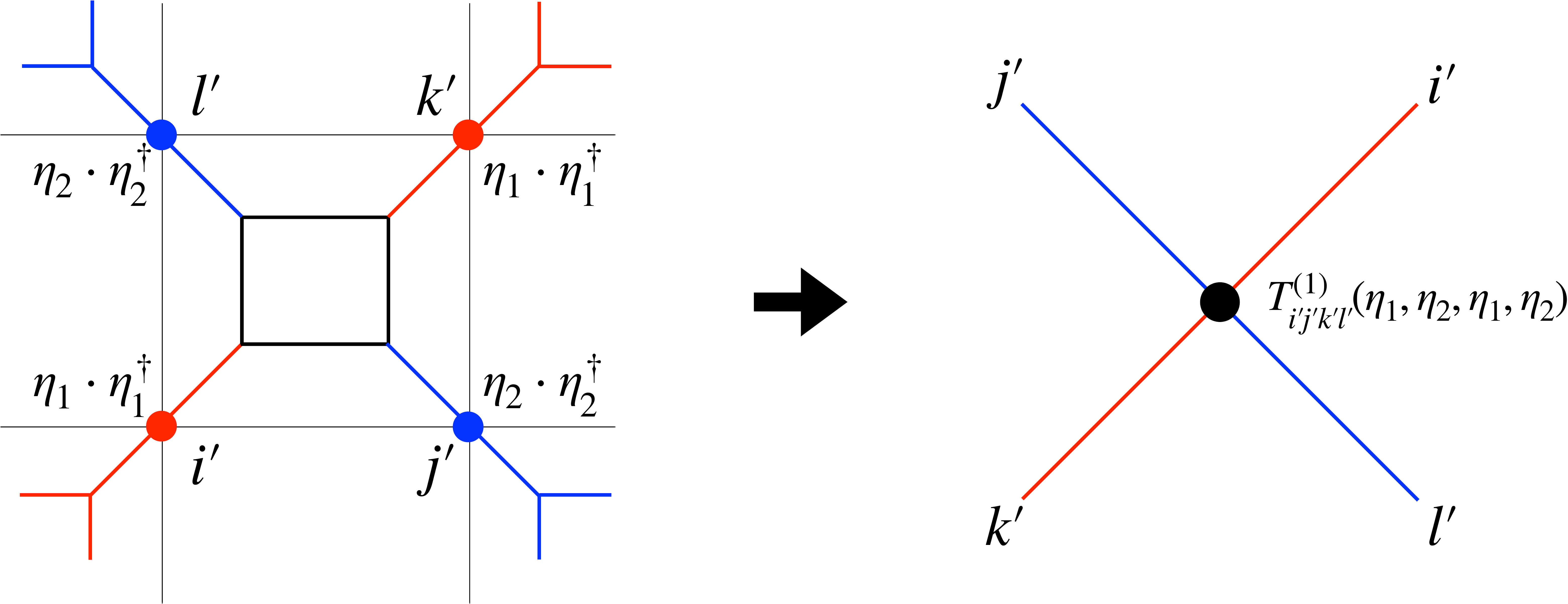} 
\end{center}
\caption{\label{fig:common} 
RG transformation of the tensor network with common noises.
}
\end{figure}

In order to preserve the lattice isotropy which is important to efficiently carry out 
multiple steps of RG transformation, 
we notice that the same noise vector should be used for obtaining $\bar{S}_{1,3}$ or $\bar{S}_{2,4}$. 
Thus, a simple way of coarse graining is to use a common noise vector for each site.
A renormalized tensor is defined as 
\begin{align}
\label{eq:renormalization}
T^{(1)[\ell]}_{i'j'k'l'}({\bm \eta}_1,{\bm \eta}_2,{\bm \eta}_1,{\bm \eta}_2) = {\rm Tr}[ \bar{S}_{1i'}({\bm \eta}^{[\ell]}_1) \bar{S}_{2j'}({\bm \eta}^{[\ell]}_2) 
\bar{S}_{3k'}({\bm \eta}^{[\ell]}_1) \bar{S}_{4l'}({\bm \eta}^{[\ell]}_2)], 
\end{align}
where the tensor indices $i'j'k'$, and $l'$ run from 1 to $D_{\rm svd}+N_r$,
and hence the bond dimension $D_{\rm cut}$ for a renormalized tensor $T^{(1)}$ is given as $D_{\rm cut}=D_{\rm svd}+N_r$.
Since the same renormalized tensor is obtained at each site, 
it is sufficient to calculate a tensor contraction at a single plaquette for each RG step, 
while ${\bm \eta}^{[\ell]}_1$ and ${\bm \eta}^{[\ell]}_2$ are different in general (See Fig.~\ref{fig:common}).
By iterating the RG process, 
we obtain a sequence of renormalized tensors for each sample $\ell$, 
\begin{align}
T^{(0)} \to T^{(1)[\ell]} \to T^{(2)[\ell]} \cdots \to T^{(n)[\ell]}, \ \ (\ell=1, \cdots, N),
\end{align}
in which the ensemble of noise vectors should be generated for each RG step independently. 
Thus,  any physical observables are calculated as an ensemble average in terms of $T^{(n)[\ell]}$. 
We note, however, that there exists an unwanted additional systematic error 
that arises from a cross contamination effect due to a multiple use of the same noise vectors. 
We will discuss the effect later, and in order to avoid the additional systematic error we will 
consider position-dependent noise vectors.

\begin{figure}[tbph]
\begin{center}
\includegraphics[clip,width=0.4\textwidth]{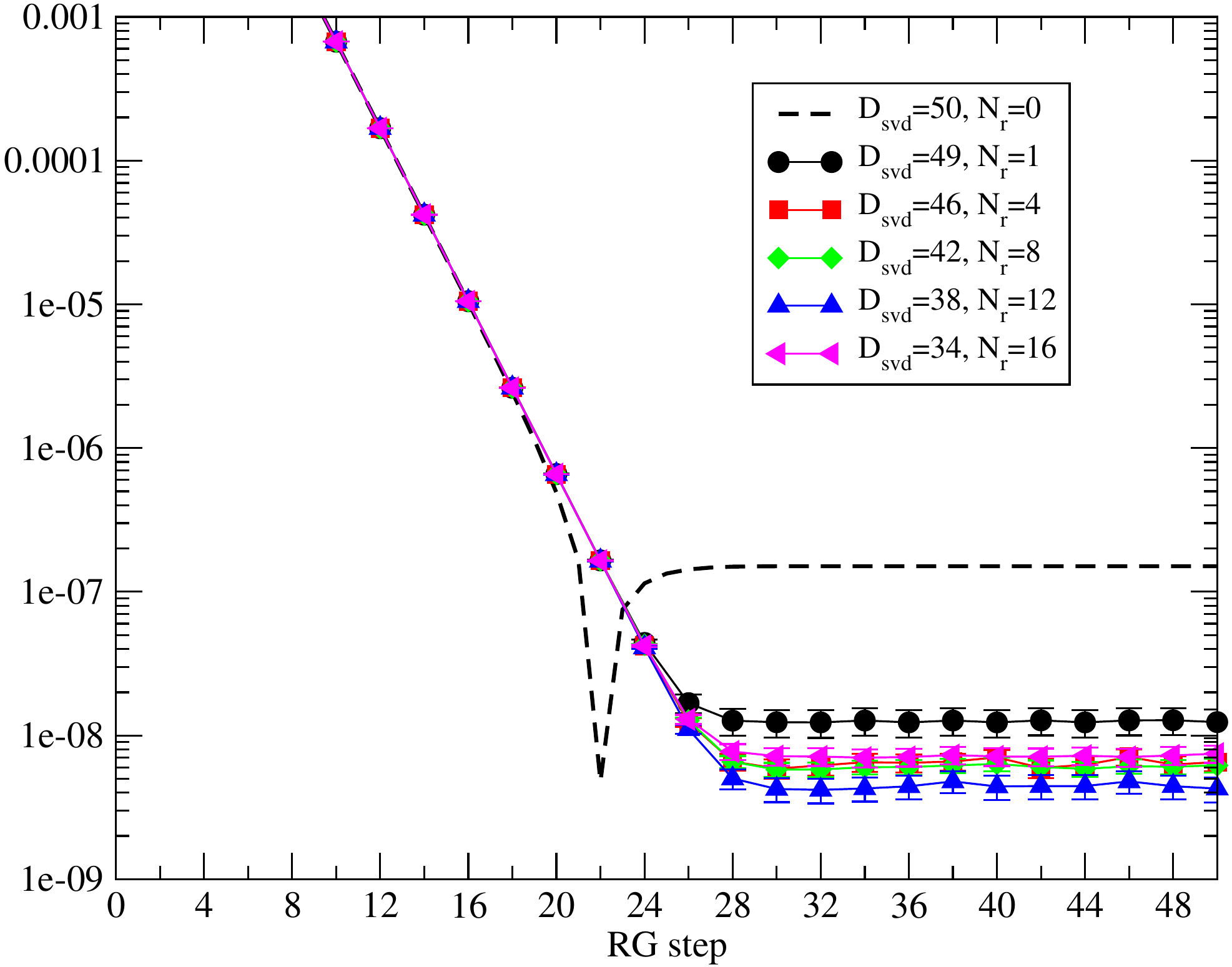} 
\ \ 
\includegraphics[clip,width=0.4\textwidth]{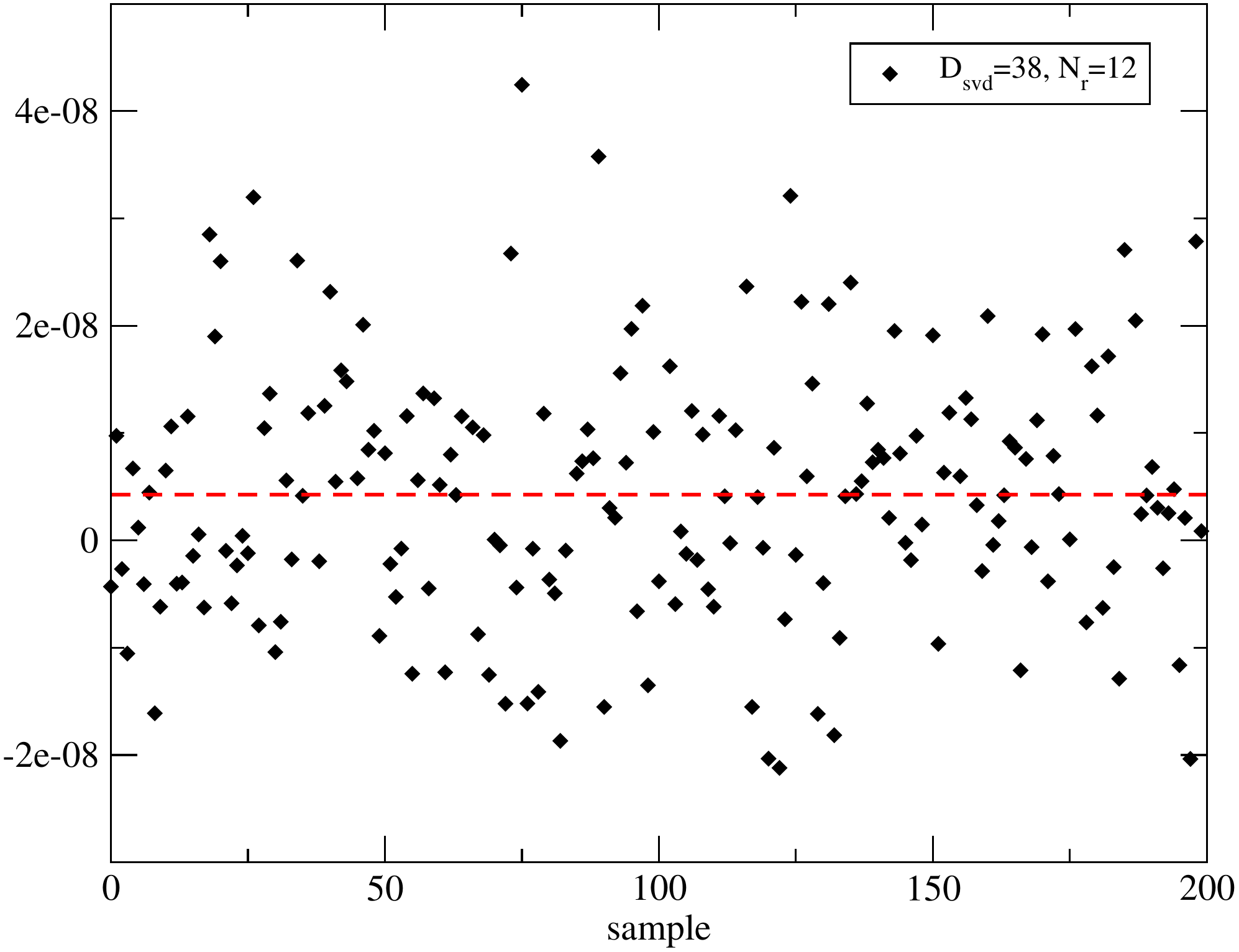} 
\end{center}
\caption{\label{fig:D50} 
Benchmark results for the RG flow in the common noise method. 
(Left) Relative deviation between our result and the exact result 
for the free energy density $|\delta f|$ with several values of the noise dimensions at the critical temperature $T=T_c$.  
The number of statistics is $N=200$ for all the results and the bond dimension is fixed to $D_{\rm cut}=50$. 
The TRG result ($D_{\rm svd}=50, N_r=0$) is also shown for comparison.
(Right) The sample dependence of $\delta f$ at $n=50$ (RG step).
The dashed (red) line represents the mean value.
}
\end{figure}

As a benchmark study of the common noise method, 
we calculate the free energy density $f$ in the 2-D Ising model. 
We employ the $Z_2$ noise, i.e., $\eta_{ir}$ takes $\pm 1$ randomly.
We calculate an averaged $f$ with $N$ statistics at the $n$-th RG step as 
\begin{align}
f = \frac{1}{2^n} \frac{1}{N} \sum_{\ell=1}^N (-T \log{Z(T^{(n)[l]})}), \quad
Z(T^{(n)[l]}) = {\rm Tr}\left[T^{(n)[l]}\right]= 
\sum_{i,j=1}^{D_{\rm{cut}}} T^{(n)[l]}_{ijij}.
\end{align}
The statistical error is estimated from the variance of the data, 
where we discard some unphysical data ($Z(T^{(n)[l]})<0$) from the analysis. 
In the left panel of Fig.~\ref{fig:D50}, we show the result of the relative deviation from the Onsager's exact result, 
$\delta f= (f-f_{\rm ex})/f_{\rm ex}$ 
at the critical temperature $T=T_{\rm c}$
for several values of noise vectors. 
In comparison, the results obtained from the original TRG ($D_{\rm svd}=50, N_r=0$) are also shown. 
We can observe a much better accuracy than the original TRG (more than $\mathcal{O}(10^{-1}$) magnitude difference) 
by optimizing the numbers of singular values and noise vectors. 
We also show a scattering plot of $\delta f$ for each sample in the right panel of Fig.~\ref{fig:D50}. 
The fluctuation is well controlled 
and we find that all the samples except the unphysical samples have better accuracy than the original TRG.
We also observe a plateau (RG fixed-point tensors)  for $n \gtrsim 28$ as in the case of the original TRG,
in which the statistical error does not grow as increasing the RG steps.
This plateau also indicates that there exists a systematic deviation that can not be reduced by increasing the statistics.
We consider this systematic deviation as a cross contamination effect due to a multiple use of the same noise vectors.
Since we use a common noise vector for each RG process, e.g. in $\bar{S}_{1,3}$ (Eq.\eqref{eq:renormalization}), 
we have a contact term originated from e.g. $\frac{1}{N_r^2} \sum_{r,s=1}^{N_r} \eta_{ir} \eta_{rj}^* \eta_{ks} \eta_{sl}^*$, 
which causes a systematic deviation from the Kronecker deltas. 
In fact, the existence of a contact term is inevitable to preserve the isotropy,
so that we could not completely remove all the systematic error in the common noise method. 
We note, however, that the resulting error is much smaller than the original TRG.
It should be also noted that this noise method can be straightforwardly implemented in other tensor networks
with the same order of the computational cost per sample.

\subsection{Coarse graining with position-dependent noise}
\begin{figure}[tbph]
\begin{center}
\includegraphics[clip,width=1\textwidth]{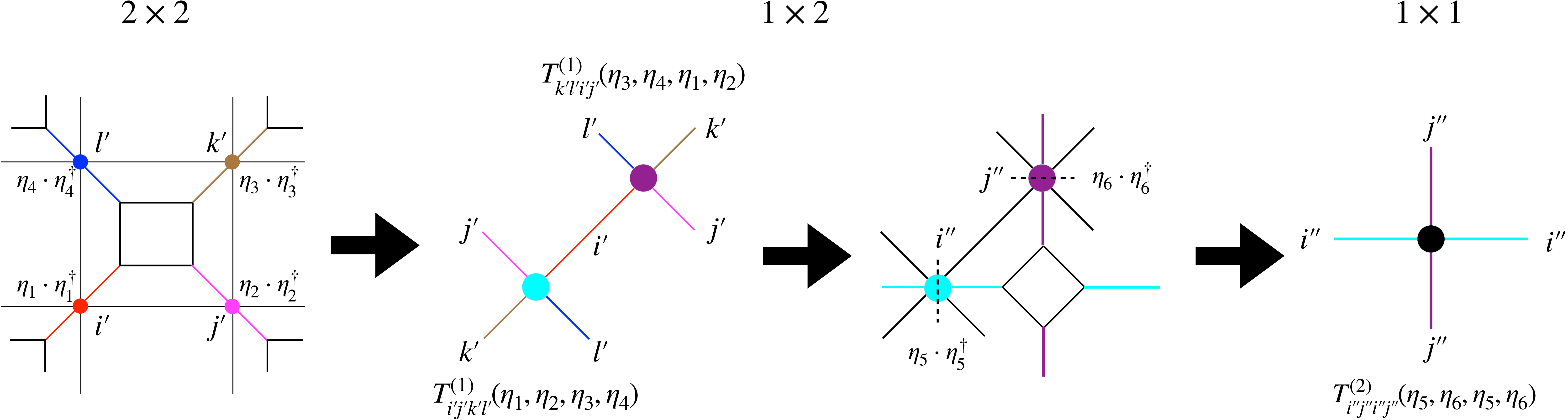} 
\end{center}
\caption{\label{fig:independent} 
RG transformation of the tensor network with position-dependent noise vectors for $2\times 2$ lattice.
In the initial lattice four different noise vectors ${\bm \eta}_{1,2,3,4}$ are 
distributed to the four sites.
The tensor contraction for each plaquette is carried out.
After the RG transformation, there exists a $1\times 2$ lattice, 
where two different noise vectors ${\bm \eta}_{5,6}$ are distributed.
The tensor contraction for one plaquette is carried out, 
and a renormalized tensor $T^{(2)}$ is obtained.}
\end{figure}

In order to completely remove the systematic error, 
we shall consider position-dependent noise vectors. 
The renormalized tensor is defined as 
\begin{align}
\label{eq:contg}
T^{(1)[\ell]}_{i'j'k'l'} = {\rm Tr}[ \bar{S}_{1i'}({\bm \eta}^{[\ell]}_1) \bar{S}_{2j'}({\bm \eta}^{[\ell]}_2) 
\bar{S}_{3k'}({\bm \eta}^{[\ell]}_3) \bar{S}_{4l'}({\bm \eta}^{[\ell]}_4)], 
\end{align}
where ${\bm \eta}^{[\ell]}_i$ is independently generated at each site $i$. 
We refer this method as position-dependent noise method.
Since the isotropy is completely broken, 
we have to calculate a renormalized tensor for each plaquette, 
where we also have to take into account the boundary conditions in a finite volume.
Therefore, the system volume $V$ has to be fixed, 
resulting in a restriction on the maximum number of the RG transformations.
See Fig.~\ref{fig:independent} for multiple steps of RG transformations in this method.
At each coarse-graining step, we obtain a statistical ensemble of tensor configurations, 
so that this approach is similar to a Monte Carlo method, 
where the probability of a spin variable is given by the Boltzmann distribution,
while in this case the tensor variables are generated 
randomly by noise vectors in analogous to a random walk. 
Thanks to this tensor configurations 
the long distance correlation can also be taken into account automatically.  
If we repeat this RG process until being a single tensor, 
the total computational cost will be proportional to the volume. 
Therefore, this method has the disadvantage of computational cost scaling as $\mathcal{O}(V D^6)$ per configuration.
It, however, guarantees that a correct result should be obtained within the statistical error, as we desired. 

\begin{figure}[tbph]
\begin{center}
\includegraphics[clip,width=0.3\textwidth]{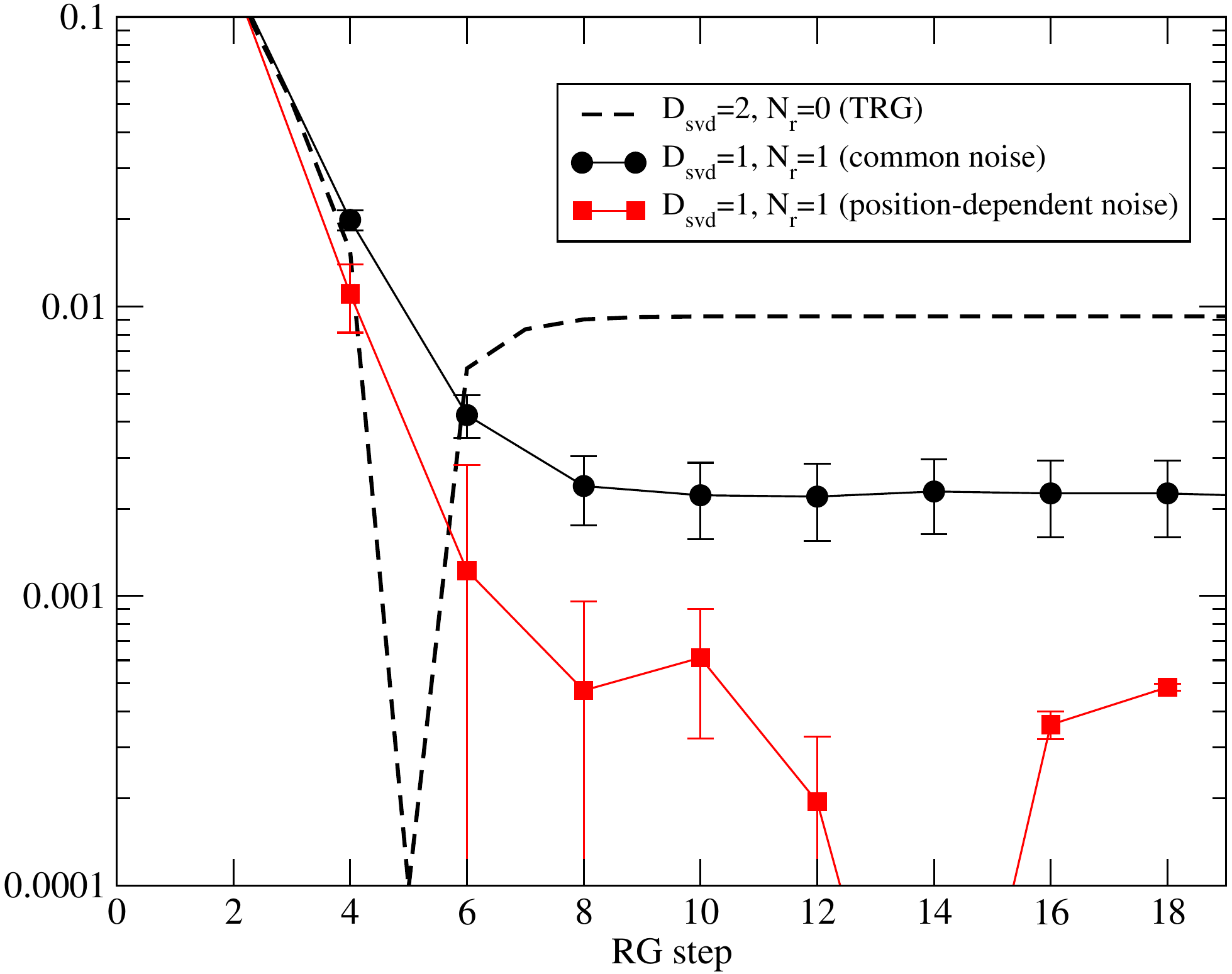} 
\end{center}
\caption{\label{fig:D1-1} 
Preliminary results for the position-dependent noise method 
in comparison with the common noise method and the original TRG for $N=20$.
}
\end{figure}

Fig.~\ref{fig:D1-1} shows our preliminary results for the position-dependent noise method.  
Since an ensemble average of the partition function $Z^{(n)}=\frac{1}{N}\sum_{\ell=1}^N T^{(n)[\ell]}$ has no systematic error, 
we calculate the free energy density from the partition function as $f=-\frac{T}{2^n}\log{Z^{(n)}}$. %
The results give a much more accurate result than 
both the common noise and the original TRG methods for each RG step. 
But the results largely fluctuate depending on RG steps and the statistical error may not be correctly estimated 
for larger RG steps due to a small number of statistics $(N=20)$.
Since the fluctuation comes from truncated components of the SVD, 
a more stable and accurate result should be obtained by increasing $D_{\rm svd}$.
A detailed study will be reported in Ref.~\cite{OT}.

\section{Summary}

We have proposed new low-rank approximation methods for TRG 
by utilizing random noise vectors in combining with the SVD.
In this method we generate tensor ensembles, 
and any physical observables are statistically calculated from this ensembles.
In the case of the common noise method, 
there exists a small systematic error due to a multiple use of the same noise vectors. 
In order to completely remove the truncation error, 
we consider position-dependent noise vectors, 
by which a statistical ensemble for the tensor configuration is generated. 
Since there is no systematic error, 
it is guaranteed that the result should become exact in the infinite statistical limit with finite $D_{\rm cut}$.
We have tested these two methods in the 2-D Ising model, 
where we obtain a better accuracy in both cases than the original TRG
with a small statistics. 
The computational costs for the common noise method per sample is the same order 
of the original TRG. 
On the other hand, the position-dependent noise method will cost as expensive as $\mathcal{O}(VD_{\rm cut}^6)$.
It should be emphasized that 
our noise methods can be implemented to other coarse graining processes of tensor networks 
as long as the SVD is used.
A more detailed study of the noise methods are given in Ref.~\cite{OT}.

\section*{Acknowledgments}
\vspace{-1mm}
We would like to thank Shinji Takeda for fruitful discussions.
M. T. is supported by U.S. DOE grant \#DE-SC0010339.
H. O. is supported in part by JSPS KAKENHI Grant Numbers 17K14309, 18H03710, and 21K03554.

\end{document}